\documentclass[epj]{svjour}

\usepackage[intlimits]{amsmath}
\usepackage{amssymb}
\usepackage{longtable}
\usepackage{multirow}
\usepackage{setspace}
\usepackage{graphicx}
\usepackage[latin1]{inputenc}
\usepackage{textcomp}         
\usepackage{array}
\usepackage{nicefrac}
\usepackage{slashed}
\usepackage[usenames,table]{xcolor}
\usepackage[colorlinks=true,linkcolor=blue,citecolor=blue,urlcolor=blue]{hyperref}
\usepackage[numbers,sort&compress]{natbib}

\begin{document}

\title{Spectra of heavy mesons in the Bethe-Salpeter approach}

\author{Christian S. Fischer\thanks{\email{christian.fischer@physik.uni-giessen.de}}
   \and Stanislav Kubrak\thanks{\email{stanislav.kubrak@theo.physik.uni-giessen.de}} 
   \and Richard Williams\thanks{\email{richard.williams@theo.physik.uni-giessen.de}}}
\institute{Institut f\"ur Theoretische Physik, Justus-Liebig--Universit\"at Giessen, 35392 Giessen, Germany.}

\date{Received: date / Revised version: date}

\abstract{
We present a calculation of the spectrum of charmonia, bottomonia and $B_c$-meson 
states with `ordinary' and exotic quantum numbers. We discuss the merits and
limitations of a rainbow-ladder truncation of Dyson-Schwinger and Bethe-Salpeter 
equations and explore the effects of different shapes of the effective running
coupling on ground and excited states in channels with quantum numbers $J \le 3$.
We furthermore discuss the status of the $X(3872)$ as a potential (excited) quark-antiquark 
state and give predictions for the masses of charmonia and bottomonia in the 
tensor channels with $J=2,3$.
}


\maketitle

\section{Introduction}

With the spectacular success of Belle, Babar, BES and the LHC 
experiments and their discovery of an ever increasing and
largely unexplained number of XYZ-states, hadron spectroscopy 
in the heavy quark region became a fascinating topic in the 
past years. Many of the newly discovered states are surprisingly 
narrow, with some of these states electrically charged and
therefore not accounted for by the conventional quark model picture 
of quark-antiquark meson bound states.
Certainly, the potential of these states to guide us in our
understanding of the underlying physics of the strong interaction
is enormous, as detailed e.g. in Refs.~\cite{Brambilla:2010cs,
Pakhlova:2010zza,JohanMesschendorpfortheBESIII:2013vla,Bodwin:2013nua} 
and references therein.

From a theoretical QCD perspective charmonia (and, to a perhaps lesser 
degree, bottomonia) are extremely interesting since they combine effects 
of non-perturbative QCD with perturbative concepts in the heavy quark 
regime. In general, the charm quark is not heavy enough to be considered 
as non-relativistic. Thus especially excited states in the charmonium spectrum 
have to be considered in a framework that is genuinely relativistic or, 
at least, incorporates relativistic corrections. Model calculations in 
terms of relativistic quasipotentials reproduce many features of the spectrum
\cite{Godfrey:1985xj,Ebert:2002pp,Ebert:2011jc,LlanesEstrada:2011kc} and provide important
guidance on the structure of the spectrum. However, in order to gain a more 
systematic understanding of the underlying physics of the strong 
interaction it is mandatory to employ approaches that are directly rooted 
within QCD. At least two different strategies have been employed in this 
direction. On the one hand, lattice gauge theory as well as non-relativistic 
QCD (NRQCD) and potential NRQCD have made substantial progress in determining 
the details of the heavy quark potential from QCD~\cite{Brambilla:2004jw,Koma:2006si}. 

On the other hand, the heavy quarkonia states can be directly calculated 
from the underlying QCD Lagrangian without the need to resort to expansions 
in terms of quark velocities or heavy quark masses. Such approaches have the
obvious advantage that the heavy and the light quark sectors can be treated 
in the same framework. In the charmonium sector, lattice gauge theory has made 
an ever increasing effort to determine the spectrum of ground and excited states
as well as exotics in dynamical calculations, see e.g. 
\cite{Namekawa:2011wt,Bali:2011rd,Liu:2012ze,Moir:2013ub,Kalinowski:2013wsa}
and references therein as well as \cite{Mohler:2012gn,Prelovsek:2013cta} for short reviews.

An alternative approach based on QCD is the relativistic functional framework 
employing the Dyson-Schwinger and Bethe-Salpeter equations. Within the
rainbow-ladder approximation first studies of the quarkonium 
spectrum~\cite{Bhagwat:2006pu,Blank:2011ha,Hilger:2014nma,Popovici:2014pha} 
as well as exotic states like tetraquarks~\cite{Heupel:2012ua} 
in the heavy quark region have been performed, accompanied by 
systematic studies in the limit of static quarks~\cite{Popovici:2010mb,Popovici:2011yz,Popovici:2014usa}.

In this work we refine and extend these calculations in two directions. On 
the one hand we include states with higher angular momentum up to $J=3$.
On the other hand we perform a systematic study of the influence of 
details in the momentum dependence of the underlying effective running
coupling on the spectrum of ground and excited states in these channels.
We extend our analysis of~\cite{Fischer:2014xha} to the heavy $\bar{c}c$, $\bar{b}b$ 
and $\bar{b}c$ mesons. In the process, we generalize the frequently used Maris-Tandy 
interaction in order to explore the impact of the shape of the interaction, with 
an emphasis on the resultant splitting between different meson channels and their
excited states. 

The paper is organized as follows. In section~\ref{sec:framework} we summarize 
the framework of the DSE and BSEs, together with a discussion of the rainbow-ladder 
interaction employed. Our results are presented and discussed in section~\ref{sec:results}. 
We conclude in section~\ref{sec:conclusions}.

\section{Framework}\label{sec:framework}
We work with the one-particle irreducible Green's functions of QCD in Euclidean space,
obtained through solutions of their corresponding Dyson-Schwinger equations (DSEs). 
With the quark propagator decomposed as
\begin{align}\label{eqn:quarkpropagator}
S^{-1}(p) = Z_f^{-1}(p^2)\left( i\slashed{p} + M(p^2)\right)\;,
\end{align}
where $Z_f(p^2)$ is the quark wave function and $M(p^2)$ its mass function, we solve
its DSE
\begin{align}\label{eqn:quarkdse}
S^{-1}(p) & = Z_2S_0^{-1}(p) \\
          & + g^2 Z_{\mathrm{1f}}C_F\int_k \gamma^\mu S(k+p)\Gamma^\nu(k+p,p)D_{\mu\nu}(k)\;.\nonumber
\end{align}
For brevity, we write $\int_k=\int d^4k/(2\pi)^4$.
The bare propagator, $S_0^{-1}(p)$ is obtained from Eq.~\eqref{eqn:quarkpropagator} by setting
$Z_f(p^2)=1$ and $M(p^2)=m_0$, with $m_0$ related to the renormalized quark mass $m_q$ 
by $m_0=Z_m\, m_q$. The renormalized coupling of QCD is denoted by $\alpha = g^2/(4\pi)$
and $Z_2$, $Z_m$ and $Z_{\mathrm{1f}}$ are the renormalisation factors of the quark
wave function the quark mass and the quark-gluon vertex. Colour traces yield the Casimir
factor $C_F=4/3$.

The non-trivial inputs into the quark DSE are the gluon propagator $D_{\mu\nu}(k)$, and the dressed 
quark-gluon vertex $\Gamma^\nu(k,p)$. Since we work in Landau gauge the gluon propagator is
transverse and given by 
\begin{align}\label{eqn:gluon}
D_{\mu\nu}(k) =T_{\mu\nu}(k)\frac{Z(k^2)}{k^2}\;.
\end{align}
where $T_{\mu\nu}(k)=\delta_{\mu\nu} - k_\mu k_\nu/k^2$ is the transverse projector.
We will discuss the quark-gluon vertex and the details of our truncation below.

Bound states of a quark and an antiquark are described by the (homogeneous) Bethe--Salpeter equation
for the corresponding Bethe--Salpeter amplitude $\Gamma(p;P)$
\begin{align}
  \left[\Gamma(p;P)\right]_{tu} = \lambda(P_i^2)\! \int_k\!\!
                                  K^{(2)}_{tr;su}(p,k;P)\left[S_+\Gamma(k;P)S_-\right]_{rs}\,,
\end{align}
with a discrete spectrum of solutions found at $P^2=-M_i^2$ for eigenvalues $\lambda\left(P_i^2\right)=1$.
The quarks $S_\pm=S(k_\pm)$ carry momentum $k_\pm = k + (\xi -1/2 \pm 1/2) P$ with momentum
partitioning $\xi$. Since the equation is manifestly covariant, all solutions are independent of $\xi$.
The quantum numbers of the bound-state under consideration follow from the tensor structure 
of $\Gamma(p;P)$. The two-particle irreducible quark anti-quark interaction kernel $K^{(2)}_{tr;su}(p,k;P)$ 
is chosen to be consistent with Eq.~\eqref{eqn:quarkdse} and the axial-vector Ward-Takahashi identity such 
that the chiral properties of the pion are preserved: the pion is both, a bound state of a quark and
an antiquark and a massless Goldstone boson in the chiral limit.

\subsection{The rainbow-ladder approximation}
In the rainbow-ladder truncation scheme one replaces the combined effects of the
dressed gluon propagator and \linebreak dressed quark-gluon vertex by a one-gluon exchange
model with effective coupling and bare vertex.
In the light quark sector the most important merit of the
rainbow-ladder scheme is its compliance with chiral symmetry such that
the (pseudo-)Goldstone boson nature of the pseudoscalar mesons and the 
associated Gell-Mann--Oakes--Renner relation are satisfied. In the opposite 
limit of very heavy quarks it has the (perhaps surprising) tendency to become 
exact~\cite{Popovici:2010mb,Popovici:2011yz,Popovici:2014usa}. For realistic masses of the charm and bottom 
quarks the static limit is relevant in the sense that potentials using the 
vector structure of one-gluon exchange only are able to reproduce global
features. Nevertheless, when it comes to the quantitative details such as the
spin-orbit splitting, sizeable corrections occur. On the other hand, it has 
been argued in~\cite{Villate:1992np} that the constraints of chiral symmetry
still play an important role in the heavy quark region. We conclude from this 
that systematic studies of the feasibility of the rainbow-ladder scheme in the
heavy quark region provide an important systematic link between the chiral and the
static limit of QCD. We will see that our study nicely complements our 
Ref.~\cite{Fischer:2014xha} where we have elucidated the assets and shortcomings of the
rainbow-ladder scheme for the spectrum of light quarks. Here we provide a similar 
analysis for heavy quarks.  

In the rainbow-ladder scheme the quark-gluon interaction appearing in the quark DSE is
combined into an effective interaction $\alpha_{\mathrm{eff}}(q^2)$
\begin{align}
  Z_{\mathrm{1f}}\,g^2 \,D_{\mu\nu}(k) \,\Gamma_\nu(k+p,p)\rightarrow 4\pi Z_2^2 \, 
                         T_{\mu\nu}(k) \,\frac{\alpha_{\mathrm{eff}}(k^2)}{k^2}\,\gamma_\nu\,,
\end{align}
with the appearance of $Z_2^2$ following from the
Slavnov-Taylor identities to maintain multiplicative renormalizability.

The corresponding symmetry-preserving two-body kernel is given by
\begin{align}\label{eqn:ladder}
  K^{(2)}_{tr;su} = 4\pi \,Z_2^2 \,\frac{\alpha_{\mathrm{eff}}(k^2)}{k^2}\,
                    T_{\mu\nu}(k)\,\left[\gamma^\mu\right]_{tr}\left[\gamma^\nu\right]_{su}\,\,.
\end{align}

One of the most frequently used examples of an effective quark-gluon interaction
is that of Maris and Tandy \cite{Maris:1999nt}. It consists of a term which
guarantees the correct ultraviolet behaviour of the quark-DSE according to
one-loop resummed perturbation theory and a term which is only active in the
infrared and supplies enough interaction strength to trigger chiral symmetry 
breaking. The interaction can be represented by
\begin{align}\label{eqn:generalmaristandy}
  \alpha_{\mathrm{eff}}(q^2)=\alpha_{\mathrm{IR}}(q^2)+\alpha_{\mathrm{UV}}(q^2)\;,
\end{align}
where
\begin{align}\label{eqn:generalmaristandy2}
  \alpha_{\mathrm{IR}}(q^2) &= \pi\eta^7\mathcal{P}(x)e^{-\eta^2x}\;,\\
  \alpha_{\mathrm{UV}}(q^2) &= \frac{2\pi\gamma_m\left(1-e^{-y}\right)}{\ln\left[e^2-1+(1+z)^2\right]}\;.
\end{align}
with momenta $x=q^2/\Lambda^2$, $y=q^2/\Lambda_t^2$, $z=q^2/\Lambda_{\mathrm{QCD}}^2$.
Here $\mathcal{P}(x)$ is a general polynomial that is equal to $x^2$ in the original
Maris-Tandy model. Below we will also discuss modifications of $\mathcal{P}(x)$.
Typically $\Lambda$ is constrained in the light quark sector to be equal to 
$0.72\,\mbox{GeV}$ by matching this scale to the known value of the pion 
leptonic decay constant. Some results for ground states are then found to 
be insensitive to the dimensionless parameter $\eta$, whereas the masses of the
excited states in general depend much more strongly on this parameter. 

In addition, this setup ignores the flavour dependence of the quark-gluon interaction.
For heavy quarks, recent results on the quark-gluon vertex suggest a considerable 
decrease in the dressing effects when the quark mass becomes heavy~\cite{Williams:2014iea}, 
a result in accordance with the general considerations from the beginning of 
this subsection. As a consequence, one would expect $\Lambda$ and $\eta$ to differ in the effective 
interaction for light and heavy hadron states. 
More generally one could expect the shape of the interaction to change, hence 
the generalization in Eq.~\eqref{eqn:generalmaristandy2} to feature $\mathcal{P}(x)$.
Below we therefore employ the polynomial form
\begin{align}\label{eqn:polinomMT}
  \mathcal{P}(x) = \sum_{i=1}^n a_i x^{i}\;.
\end{align}
and investigate its impact on the heavy meson spectrum restricting ourselves 
to terms with $n \le 4$.

\begin{figure*}[t]
  \begin{center}
  \includegraphics[width=0.90\textwidth]{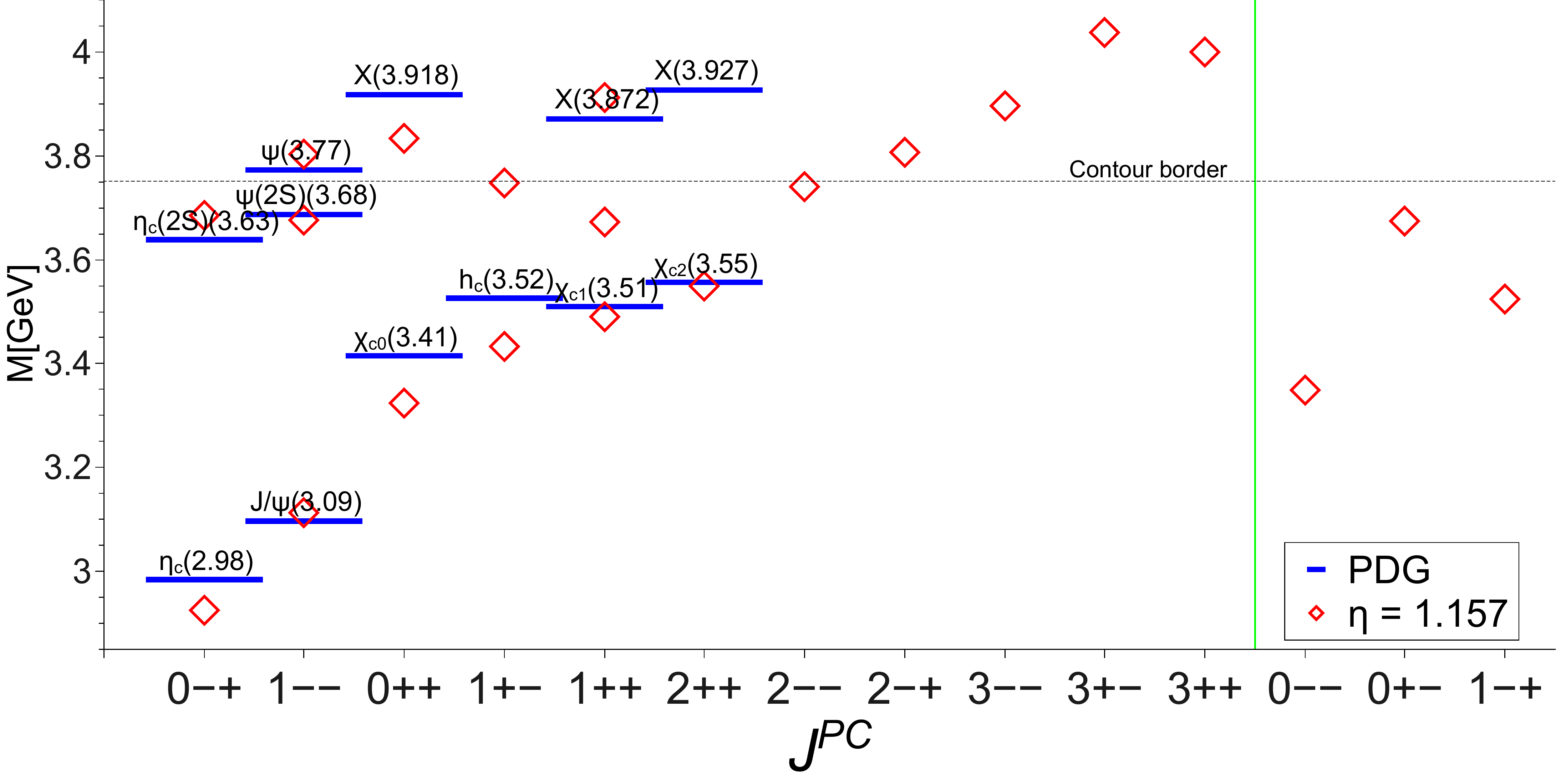}
  \caption{Spectrum of ground and excited charmonium states for the vanilla MT-rainbow-ladder interaction.
  The three rightmost states are exotic in the quark-model.}
  \label{fig:charm}
  \end{center}
\end{figure*}

\subsection{Numerical methods}\label{sec:numericalmethods}
Our numerical methods have been explained in Ref.~\cite{Fischer:2014xha} and
we refer the reader to this work for details. For the quark-DSE we use a version 
where the complex momentum is flowing through the internal quark propagator, leaving
the momentum in the Maris-Tandy interaction real. This way we avoid the sizable 
numerical errors that may occur when evaluating the effective interaction in the 
complex plane. Furthermore we use a Pauli-Villars type regulator to avoid cut-off 
effects in the quark-DSE.

In general, when evaluating bound state in channels with large angular momentum $J$ 
or radially excited states the problem arises that the Bethe-Salpeter equation evaluates
the internal quark propagators at large time-like momenta. With increasing mass of 
the state in question at some point one probes the analytic structure of the quark 
propagators, which in rainbow-ladder approximation is given by pairs of complex 
conjugate poles~\cite{Alkofer:2003jj,Fischer:2008sp,Dorkin:2013rsa}. Since the numerical 
treatment of the quark-DSE at and beyond these poles is extremely hard we refrain from 
a brute force treatment of the problem and resort to the extrapolation techniques for 
the eigenvalue of the BSE described in Ref.~\cite{Fischer:2014xha}. There we compared
two different extrapolation techniques and estimated thereby the associated error.
For ground states with masses not too far beyond the calculable domain, these errors
are generically on the level of 1 \%. For excited states and states with larger masses 
the accumulated error of extrapolation is larger. We accepted extrapolations
up to errors on the 5 \% level for the exploratory study presented here.

\section{Results}\label{sec:results}

\subsection{Charmonia}\label{sec:charm}

\subsubsection{Vanilla Maris-Tandy}\label{sec:MT}
We start our study with what we term the vanilla Maris-Tandy interaction,
i.e. we keep the scale $\Lambda=0.72$ GeV from the light meson sector and
explore the dependence of the spectrum on $\eta$. Furthermore, for the
polynomial $\mathcal{P}(x)$ in Eq.~(\ref{eqn:generalmaristandy2}) we 
use the Maris-Tandy form ($a_2 = 1$, remaining $a_i =0$). 
This original form of the Maris-Tandy (MT) interaction has been employed in the
heavy meson sector already in Ref.~\cite{Blank:2011ha}. It therefore
provides a convenient starting point. In the subsections \ref{int1}, \ref{int2} 
and \ref{int3} below, we will also discuss deviations from this form in $a_1$, 
$a_4$ or the scale $\Lambda$. 

In Ref.~\cite{Blank:2011ha} a general fit has been performed of ground state
masses in the vanilla MT-model to experimental values. Here we employ a different 
strategy. We utilise an observation made in~\cite{Fischer:2014xha}, namely that 
the channels $1^{--}, 2^{++},3^{--}$ etc. are particularly well represented in the 
rainbow-ladder framework. Within the realms of potential quark models these
states share the property that the spin-spin tensor forces do not play an 
important role. Since these states are well represented in the vanilla
MT-interaction in the light meson sector, we first concentrate
on the ground and first excited state in the $1^{--}$ channel ($J/\Psi$, $\Psi(2s)$) 
and the ground state in the $2^{++}$-channel ($\chi_{c2}$). We minimize the
deviations of our calculated masses with the experimental values under variation 
of the charm quark mass and the $\eta$-parameter in the MT-interaction. 
We obtained good agreement with experiment using a charm quark mass of 
$m(19 \,\mbox{GeV}) = 0.870 \,\mbox{GeV}$ and a value $\eta=1.157$.

Our results for all presently available channels are shown in Fig.~\ref{fig:charm},
the explicit values are all collected in Tab.~\ref{tab:results} at the end of the results section.
Since we have fixed the two input parameters from $J/\Psi$, $\Psi(2s)$ and $\chi_{c2}$, 
all other states can be viewed as model predictions. In the pseudoscalar channel
we find a mass of the $\eta_c$ which is slightly too low, but still within 3 \% of 
the experimental value. In the language of potential models, this may indicate an 
overestimation of the spin-spin contact term in the effective interaction. Very good 
agreement with experiment is obtained for the ground state in the $1^{++}$-channel, 
whereas the masses of
the scalar $0^{++}$ and the axialvector $1^{+-}$ ground states are further 
off but still within five percent of the experimental value. Similar results have
been obtained already in Ref.~\cite{Blank:2011ha,Hilger:2014nma}. The new element here is the
calculation of states with $J=3$ and the excited states. In Ref.~\cite{Fischer:2014xha}
we already observed in the light quark sector, that the rainbow-ladder interaction
is well suited to reproduce states in the sequence $1^{--}, 2^{++}, 3^{--},...$, which
are located on the same Regge-trajectory. Since we reproduce the experimental results for
the $J/\Psi$ and the $\chi_{c2}$ with an error below 1~\%, we therefore expect our 
result for the mass of the $3^{--}$-state of 
\begin{align}
  m_{3^{--}}= 3896 \,\mbox{GeV}
\end{align}
to be also accurate on this level due to uncertainties in the interaction alone.
Since this state is a ground state beyond but still close to the boundary of calculable states 
(the dashed line in the plot) it is not subject to a large extrapolation error
(see \cite{Fischer:2014xha} for a discussion of the extrapolation procedure). We 
therefore expect our prediction for the mass of this state to be quite robust, with 
a guesstimate of the overall error on the 3~\% level. Within these errors, we agree with the quark model 
prediction \cite{Ebert:2011jc} and the lattice QCD results \cite{Bali:2011rd,Liu:2012ze}.
For the other tensor ground states with $J=2$ and $J=3$ we expect our results to be 
much less accurate, with a guesstimate of total systematic errors on the 5-10~\% level. 

Similar to the light quark sector \cite{Fischer:2014xha} we also find, that the 
sequence $1^{--}, 2^{++}, 3^{--}$ lies on a Regge-trajectory with an accuracy that is
even better than in the potential model of Ref.~\cite{Ebert:2011jc}. 
For $J = \alpha M^2 + \alpha_0$ we find $\alpha = 0.36$ and $\alpha_0 = -2.55$, which 
is also somewhat steeper than the result of \cite{Ebert:2011jc}. For the heavy quark 
sector this confirms a result found in Ref.~\cite{Fischer:2014xha} for light quarks,
that Regge-type behaviour in the spectrum may be found without any direct connection 
to an underlying string-picture.

We also calculated the masses of ground states with exotic quantum numbers that cannot 
be accounted for as $q\bar{q}$-states in quark models. Our results are displayed in 
Fig.~\ref{fig:charm}. Note that within a genuinely relativistic framework such as 
lattice QCD or the functional approach used here, there is no problem representing these 
states with bilinear operators. Therefore they naturally appear also in the 
$q\bar{q}$-spectrum. Of course, it is then an open question, whether sizeable admixtures 
from states with a different quark content than $q\bar{q}$ (captured only in appropriate 
extensions of the quark-gluon interaction beyond rainbow-ladder) do exist. Furthermore,
even within the $q\bar{q}$-picture large corrections beyond rainbow-ladder may occur.
Because of these possibilities our calculated masses should be regarded with a lot of 
caution. 

For the excited states we observe very good agreement in the vector channel: our
value for the mass of the $\Psi(2S)$ is very close to the experimental one, and even
the next radial excitation is nicely represented. In the pseudoscalar channel the 
splitting between the ground and the excited state is slightly too large, making the 
agreement of the $(2S)$-state with experiment even better than for the ground state
$\eta_c$. It is interesting to observe that the resulting fine structure splitting of 
the ground and excited states show a qualitatively difference when compared with 
experiment: whereas the ground state splitting is too large the splitting in the 
excited state is too low. Such an uncorrelated behaviour of the two splittings has
also been observed in lattice QCD \cite{Bali:2011rd}.

In the `good' tensor channel $2^{++}$ potential radially excited states like
the $X(3927)$ are not reproduced in our framework. There is a considerable uncertainly 
due to the extrapolation procedure needed in this mass region (see the discussion in 
section \ref{sec:numericalmethods}), which is enhanced for excited states. 
Taking our result at face value, however, the current model would disregard 
the notion of the $X(3927)$ to be an ordinary meson state.

From an experimental point of view, the $1^{++}$-channel is perhaps the most interesting
one. There the famous $X(3872)$-state awaits its identification as a meson-molecule,
a tetraquark, or an ordinary quark-antiquark bound state. The literature on this
subject is enormous, therefore we point the reader only to Ref.~\cite{Bodwin:2013nua} 
for a first overview. The interesting question in this context is whether a description
on a quark-antiquark basis is possible at all for the $X(3872)$. In the present
rainbow-ladder model we find a ground state in this channel that is only slightly below
the experimental state $\chi_{c1}$. In addition, we find an excited state at 
$m = 3672 \,\mbox{MeV}$ that cannot be accounted for by experiment. A second excitation 
is found at $m = 3912 \,\mbox{MeV}$, close to the quark model prediction for the first
radial excitation. Indeed, we verified by inspection that the Bethe-Salpeter wavefunction
of our second excitation corresponds to the first radial excitation, have one zero
crossing at finite relative momentum between the quark-antiquark constituents. Thus we 
agree with the quark-model result, that the splitting between ground state and first 
radial excitation in the $1^{++}$-channel is too large to account for the $X(3872)$
to be a pure radially excited quark-antiquark state. This is true for the vanilla
MT-interaction, but we will argue below that modifications of the interaction within
the rainbow-ladder framework do not change this situation. What remains to be clarified
is the nature of the extra state that we see at $m = 3672 \,\mbox{MeV}$. We verified
that the leading part of its wave function has no zero crossing, thus ruling out its
interpretation as a radial excitation. It has well-defined charge conjugation and parity 
properties, but it cannot be accounted for by the naive quark model picture. 

Currently, we do not have a good explanation for the appearance of this state. In principle, 
it may or may not be that this state is spurious in the sense that it only appears in the rainbow-ladder
framework and disappears when corrections beyond rainbow-ladder are taken into account.
Clearly, the present form of the rainbow-ladder interaction is not sufficient to describe
all structures of the experimental spectrum. This is especially apparent in the $0^{++}$
and $1^{+-}$-channels. We therefore expect sizeable corrections when interactions beyond the 
rainbow-ladder approximation are taken into account\footnote{In addition, for states above 
the $D\bar{D}$-threshold coupled channel effects may play an important role.}.
This is the subject of future work. 
Here, as a first
step in this direction, we would like to explore the extent to which the rainbow-ladder 
interaction can be modified to improve the agreement with experiment. 
To this end we systematically explore the variations of the spectrum once we go away 
from the simple shape of the effective coupling Eq.~(\ref{eqn:generalmaristandy}) with 
polynomial $\mathcal{P}(x) = x^2$, i.e. $a_2=1$ and all other $a_i=0$ in 
Eq.~(\ref{eqn:polinomMT}). This is the subject of the next two subsections.

\subsubsection{Effective interaction including $a_1$}\label{int1}
\begin{figure}[!t]
  \begin{center}
    \includegraphics[width=0.38\textwidth]{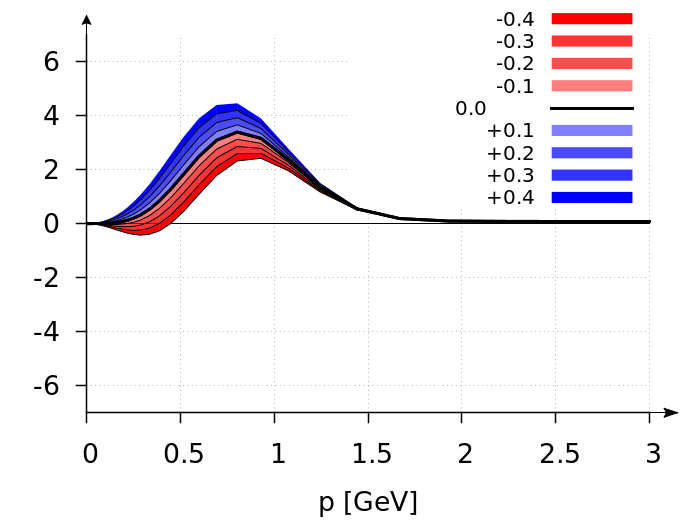}
    \caption{The shape of the effective coupling for the generalized Maris-Tandy interaction 
             with varying $a_1$ and $a_2=1$ held constant (see text for further explanations).}
    \label{fig:slope_a1}
  \end{center}
  \begin{center}
    \includegraphics[width=0.38\textwidth]{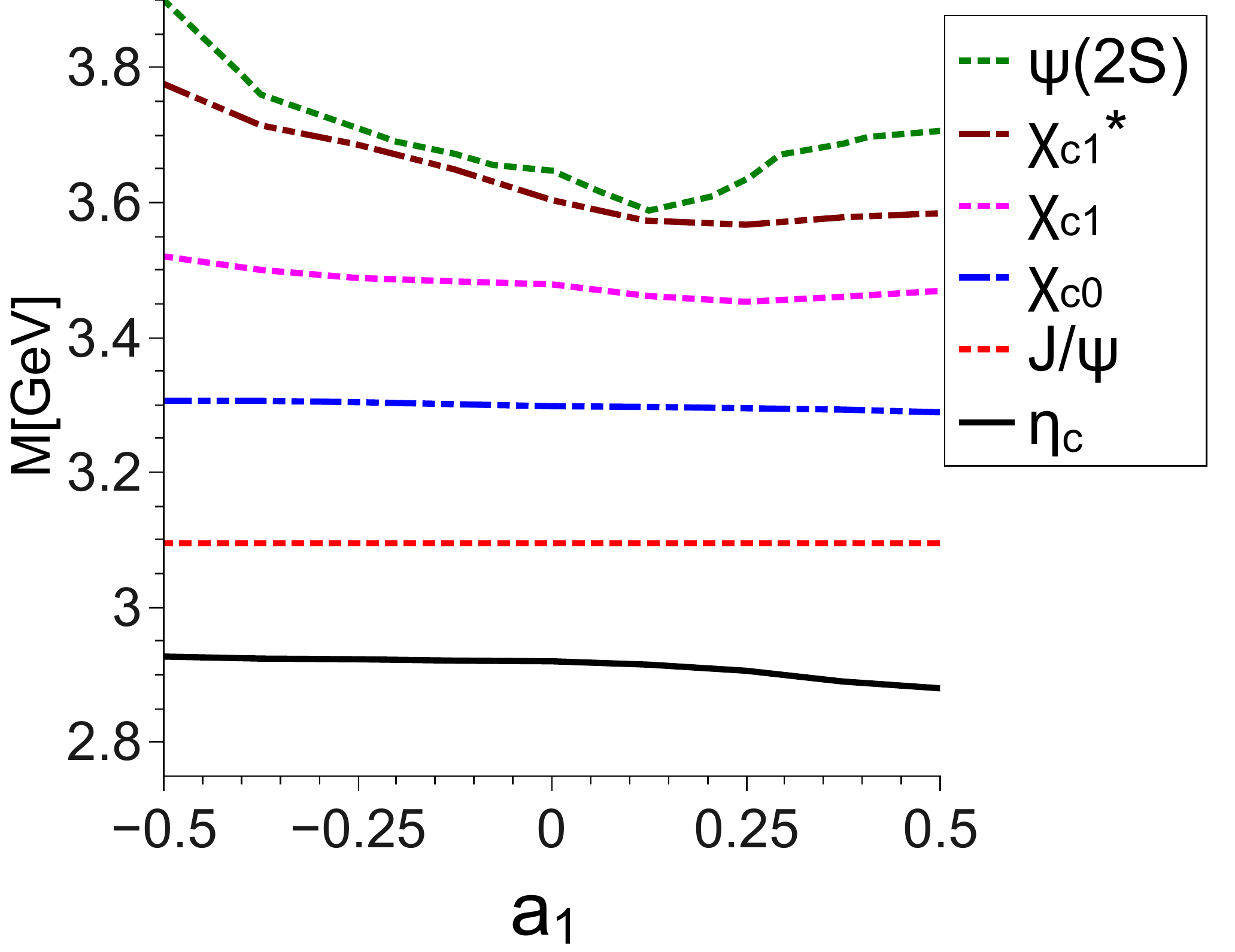}
    \caption{The response of masses of bound and excited states on the variation of the shape 
             of the effective interaction with $a_1$.}
    \label{fig:trend_a1}
  \end{center}
\end{figure}
\begin{figure}[!t]
  \begin{center}
    \includegraphics[width=0.38\textwidth]{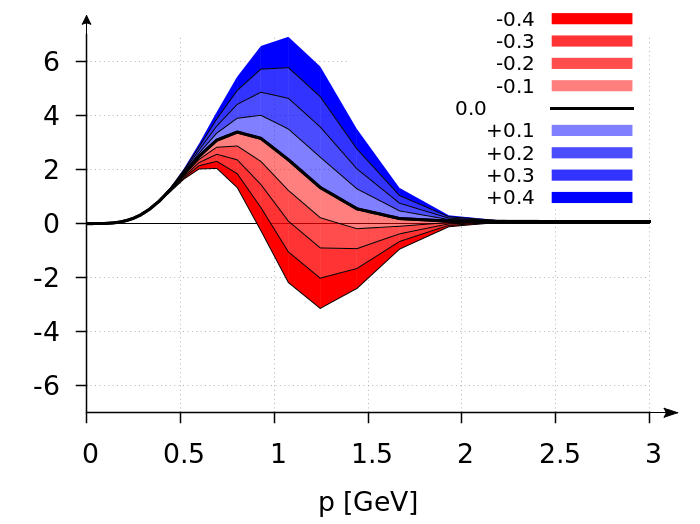}
    \caption{The shape of the running coupling for the generalized Maris-Tandy interaction with $a_2=1$, $a_1=a_3=0$ and varying $a_4$.}\label{fig:slope_a4}
  \end{center}
  \begin{center}
    \includegraphics[width=0.38\textwidth]{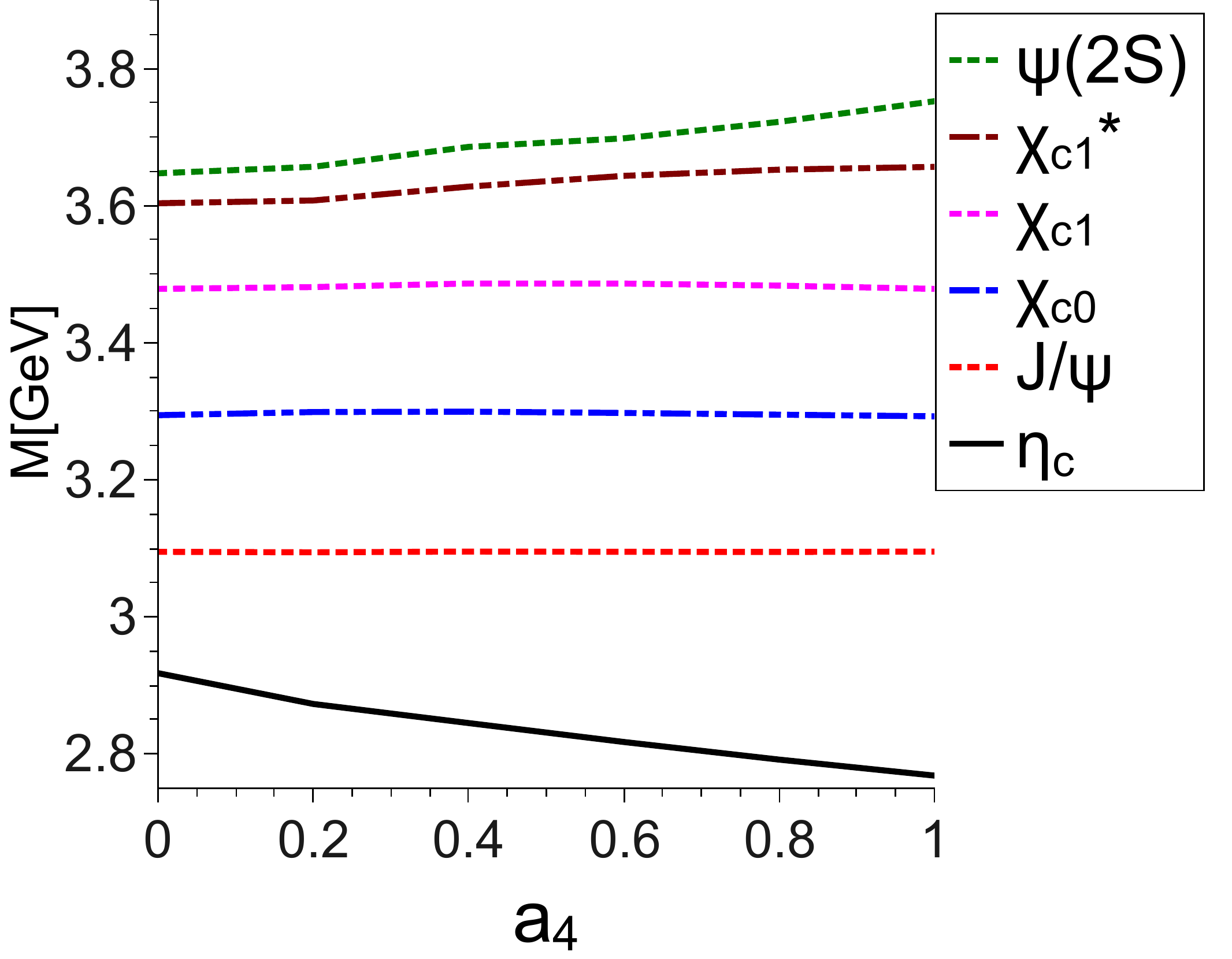} 
    \caption{The response of masses of bound and excited states on the variation of the shape of the effective interaction with $a_4$.}\label{fig:trend_a4}
  \end{center}
\end{figure}

In order to study the variations of the charmonium spectrum with respect to changes in the
general momentum behaviour of the effective coupling we now introduce
additional structure in the polynomial $\mathcal{P}(x)$ in Eq.~(\ref{eqn:generalmaristandy2}). 
First we vary $a_1$ in the interval
$-0.5 \le a_1 \le 0.5$. For the effective running coupling the resulting variation
is shown in Fig.~(\ref{fig:slope_a1}). Clearly, the integrated strength, but
also the fine details of the coupling change: For negative $a_1$ we even obtain 
a zero crossing with the corresponding scale associated with the relative
strength between the $a_1$ and $a_2$-terms (here we keep $a_2=1$). Such an 
effective coupling is unusual, but not unreasonable. Recent calculations of 
the three-gluon vertex \cite{Aguilar:2013vaa,Blum:2014gna,Eichmann:2014xya} suggest that the interplay between 
ghost and gluon degrees of freedom in the corresponding Dyson-Schwinger equation 
for the vertex may very well introduce such a zero crossing. This possibility is also 
seen in corresponding lattice calculations \cite{Cucchieri:2008qm}. Since the three-gluon 
vertex is an integral part of the non-Abelian diagrams in the DSE for the quark-gluon vertex, 
this behaviour may translate into a corresponding zero crossing of the quark-gluon 
vertex~\cite{Williams:2014iea} and subsequently into the effective coupling.

The resulting changes in the meson spectrum are displayed in 
Fig.~\ref{fig:trend_a1}. Adjusting the bare charm quark mass via $m_{J/\Psi}$
to accommodate for the changes in the integrated interaction strength we
observe only very small changes in the resulting masses for the ground state mesons.
However, the excited states turn out to be sensitive to the details of the interaction.
This is particularly true for the $\Psi(2S)$ and the first excitation in the
$1^{++}$-channel, which we denoted by $\chi_{c1}^{*}$ in order to distinguish it 
from the second excited state in this channel which we identified with the first
radial excitation $\chi_{c1}^{'}$ as discussed above. In particular for negative values 
of $a_1$, corresponding to the zero crossing of the interaction discussed above, we find much
increased values for the mass of the $\chi_{c1}^{*}$, which eventually may even  
hit the experimentally observed mass of the $X(3872)$. However, this comes at a price: 
the mass of the $\Psi(2S)$ reacts in a similar way and substantially moves away from 
the experimental value, almost reproduced for $a_1=0$. In general we find that 
variations of the infrared behaviour of our interaction via changes in $a_1$ do not 
improve the agreement of the calculated spectrum with the experimental one.

\subsubsection{Effective interaction including $a_4$}\label{int2}

Next we consider the generalized Maris-Tandy interaction, Eq.~(\ref{eqn:generalmaristandy2}), 
given by $a_1=0$, $a_2=1$ but non-trivial components $a_3$ or $a_4$. Both of these modify 
the interaction in the intermediate momentum region, while keeping the infrared and
ultraviolet behaviour untouched as can be seen from Fig.~\ref{fig:slope_a4} for the 
example of variations in $a_4$. Since variations of $a_3$ act similarly on the effective
coupling we keep $a_3=0$ fixed and restrict ourselves to variations of $a_4$.
Furthermore, we keep $a_4 \ge 0$, since there are no indications that the dressing of the
quark-gluon vertex can induce a negative effective interaction in the mid-momentum region
(in contrast to the infrared momentum region discussed in section \ref{int1} above).

Again, we study the variation of the charmonium spectrum while still readjusting
the charm quark mass to reproduce the vector ground state $J/\Psi$. Our results
are given in Fig.~\ref{fig:trend_a4}. Here we find a substantial increase in the
mass splitting between the pseudoscalar and the vector channel due to the additional
interaction strength in the mid-momentum region. At the same time, the masses of the
excited states $\Psi(2S)$ and $\chi_{c1}^{*}$ increase slightly. This moderate increase 
is, however, nowhere large enough to bring the $\chi_{c1}^{*}$ close to the observed 
$X(3872)$-state. In general we find that variations of the mid-momentum behaviour of 
our interaction via changes in $a_4$ do not improve the agreement of the calculated 
spectrum with the experimental one. 

\subsubsection{Vanilla Maris-Tandy with variation of $\Lambda$}\label{int3}
\begin{figure}[!t]
  \begin{center}
    \includegraphics[width=0.38\textwidth]{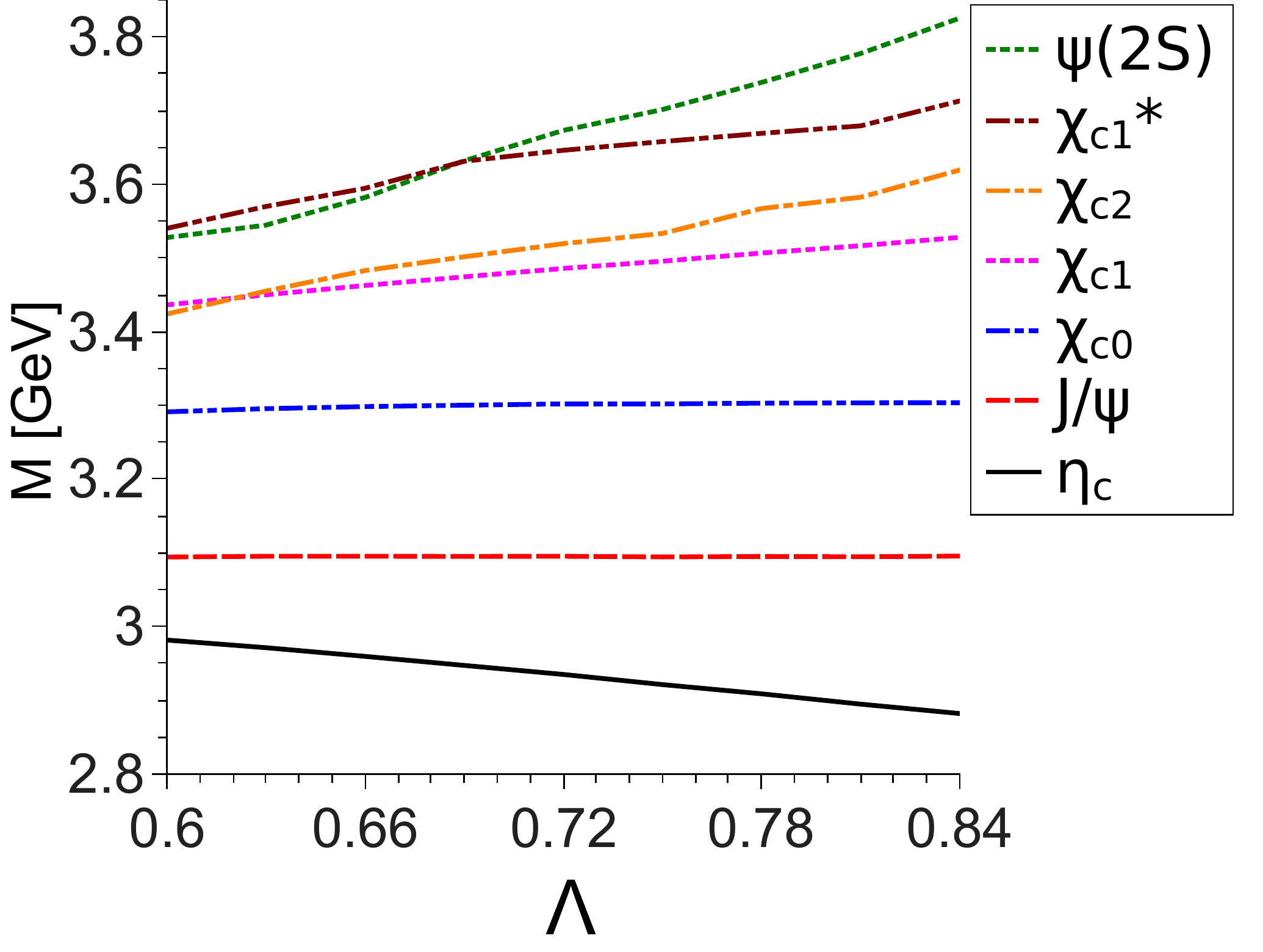}
    \caption{The response of masses of bound and excited states on the variation of the 
             scale parameter $\Lambda$ in the interaction.}
    \label{fig:trend_lambda}
  \end{center}
\end{figure}

Finally we studied the variation of the masses of the ground and excited states with
a change of the scale $\Lambda$ in the interaction. We varied the scale between 
$0.6 \,\mbox{GeV} \le \Lambda \le 0.84 \,\mbox{GeV}$, keeping $\eta = 1.157$ fixed but 
readjusting the charm quark mass such that the ground state in the vector channel
does not change. Our results are shown in Fig.~\ref{fig:trend_lambda}. We find a 
substantial variation in particular of the mass of the first excited state in the
vector channel with change of $\Lambda$. This variation is so steep, that the agreement
with the measured mass is only good in the vicinity of the scale $\Lambda = 0.72$ GeV
inherited from the light meson sector in the first place. This justifies our original choice
in section \ref{sec:MT}. 

\subsection{Bottomonia}\label{sec:bottom}

\begin{figure*}[t!]
  \begin{center}
    \includegraphics[width=0.90\textwidth]{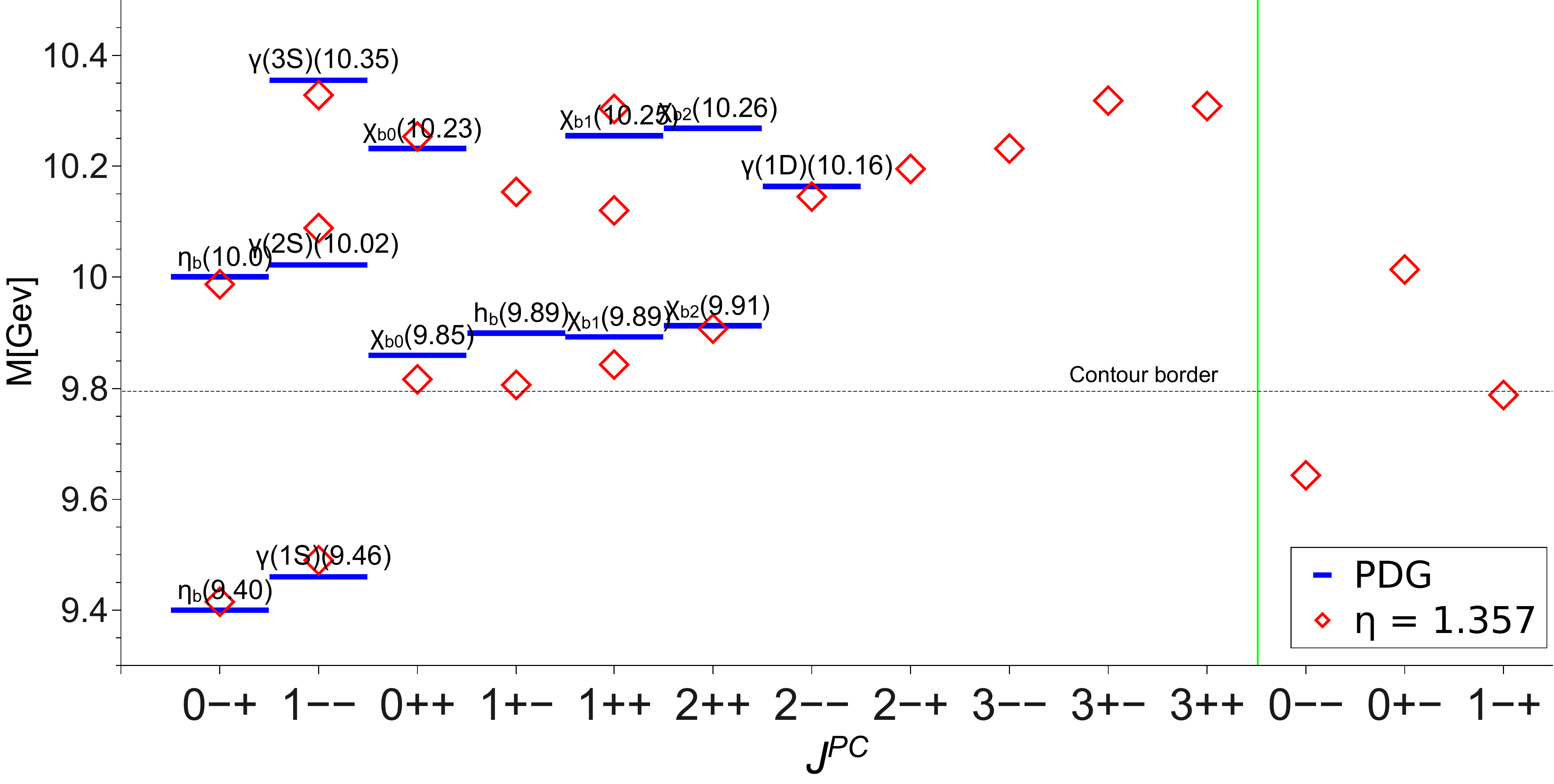}
    \caption{Spectrum of ground and excited bottomonium states for the vanilla MT-rainbow-ladder interaction.
    The three rightmost states are exotic in the quark-model.}\label{fig:bottom}
  \end{center}
\end{figure*}

Our results for the spectrum of bottomonia are shown in Fig.~\ref{fig:bottom}. Compared to
the charmonium spectrum in Fig.~\ref{fig:charm} we had to change the shape of the interaction
by adjusting the $\eta$-parameter from $\eta=1.157$ for the charm-case to $\eta=1.357$ 
for the bottom quarks. This reflects part of the underlying flavour dependence of the quark-gluon 
interaction as noted in Ref.~\cite{Williams:2014iea}. Our corresponding mass of the bottom quark
is $m(19 \,\mbox{GeV})=3.790 \,\mbox{GeV}$. The resulting spectrum of ground and excited states, 
however, has similar features when compared with experimental values as the charmonium one. 
Once again, the $0^{-+}$, $1^{--}$ and $2^{++}$ ground states are well represented. The necessary
extrapolation needed for the $2^{++}$ is still under control, since the state is not too far above
the limit where everything can be calculated (the dashed line in the plot). Surprisingly good is
also the negative parity tensor state, although the extrapolation procedure in this
mass region must be considered with a little more caution. The ground states
in the scalar and axial vector channels are further off their experimental counterparts,
although still within the 1 \% deviation margin. Thus overall, the ground state spectrum
of bottomonia is well represented in the rainbow-ladder approximation of the BSEs.
Provided the good agreement in the $2^{--}$-channel can be seen as an indication that 
extrapolation even in this mass region works well, we can regard the masses of the 
further tensor states with $J=2$ and $J=3$ as more or less solid predictions 
on the level of one percent. Compared to the quark-model predictions of \cite{Ebert:2011jc}
we find only slight deviations of the order of 30-70 MeV for the $2^{-+}$ and the states 
with $J=3$.

In contrast to the charm-case, the lowest lying excited states in the bottomonium spectrum are
already in a mass region where we need to extrapolate the eigenvalue of the BSE, as discussed
above. Nevertheless, the extrapolation procedure seems to work and the results are surprisingly
good and comparable with the corresponding ones in the charmonium spectrum, where much less
extrapolation was needed. The first excited states in the pseudoscalar, vector and even the
scalar channel are quite accurate and even the $\Psi(3S)$ works reasonably well. In the $1^{++}$-channel
we make the same observation as in the charmonium spectrum: there is a first excited state 
which is not a radial excitation, whereas the second excited state can be identified
with the first radial excitation in the channel, i.e. the $\chi^\prime_{b1}$. Again, it will be
interesting to study corrections beyond the rainbow-ladder framework.

Our results for exotic states are also given in the plot, although, as already mentioned 
for the charmonia spectrum, they should be regarded with some caution due to potential
mixing effects with non-$q\bar{q}$-states in these channels.

\subsection{Charm-bottom bound states}\label{sec:charmbottom}
\begin{figure}[!t]
  \begin{center}
    \includegraphics[width=0.38\textwidth]{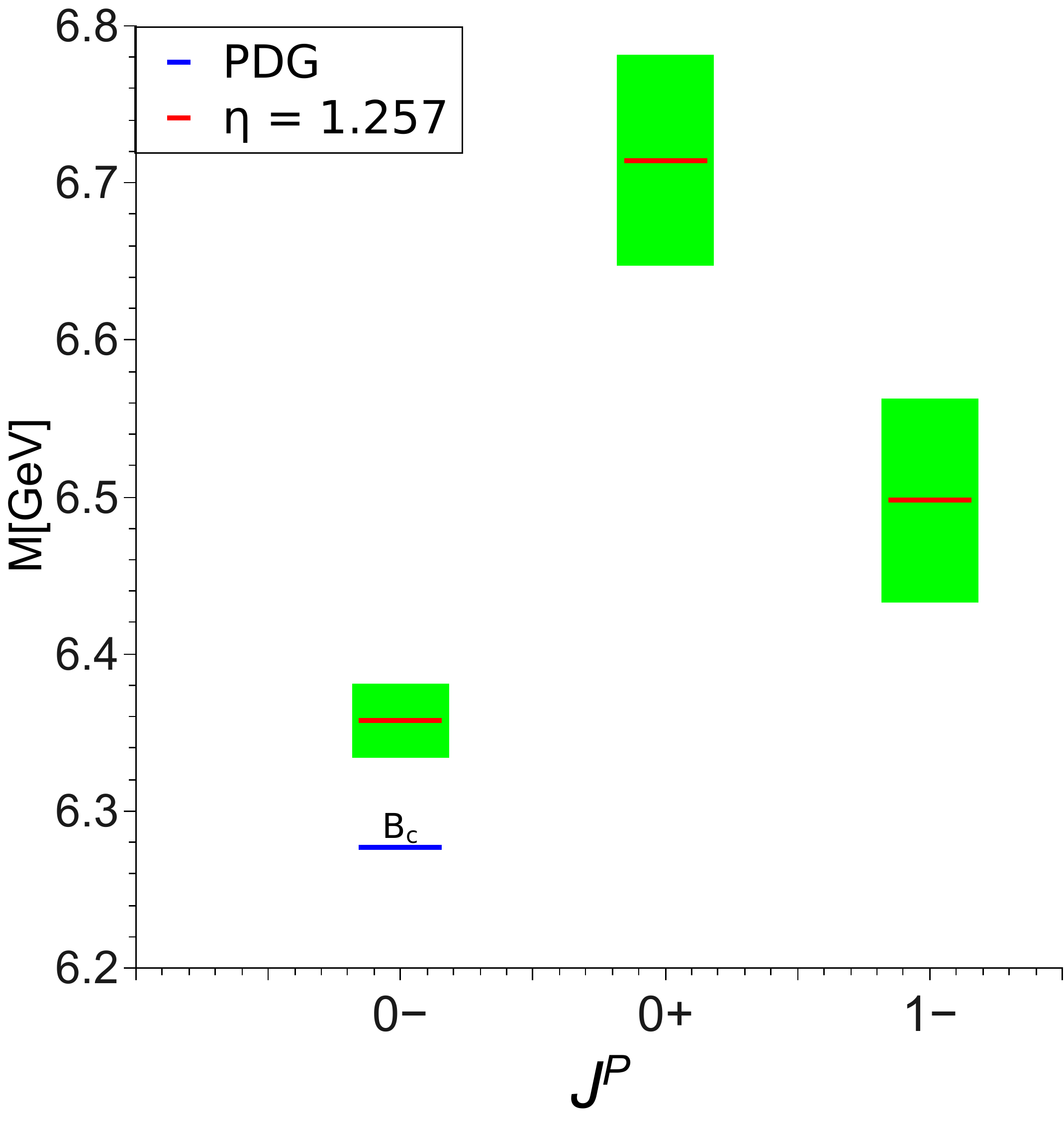}
    \caption{The calculated $b\bar{c}$ spectrum compared to experiment. The green bands correspond 
             to the variation $\eta=1.257\pm0.1$.}\label{fig:spectrumbc}
  \end{center}
\end{figure}

Finally, we present our results for selected channels of $B_c$-mesons. 
Heavy-light systems in the Bethe-Salpeter
approach are notoriously difficult to treat, since the problem of probing the analytical
structure of the internal quark propagators already appears for ground states, see e.g.
Ref.~\cite{Rojas:2014aka,Gomez-Rocha:2014vsa} for recent studies of the problem. Our results for these states,
shown in Fig.~\ref{fig:spectrumbc} are therefore all extrapolated and have a systematic error
of about 5-10 \%. In the plot we show values obtained using a variation of the $\eta$-parameter
in the interaction ranging approximately between the ones used for the charmonia and bottomonia.
In this way we heuristically take into account the varying strength of the interaction for the
two different quark flavours involved. The central value, given by the red line, corresponds 
to $\eta=1.257$. Given the inherent uncertainties in the calculation, our value for the $B_c$
in the pseudoscalar channel is surprisingly close to the experimental one. Since this is the
state with the lowest mass, the extrapolation error is also smallest. Since the rainbow-ladder
approach works well in the vector channel we consider the existence and to some extent also the 
mass of the vector state as a prediction of the approach, whereas the scalar channel has to
be considered with much more reservation. Despite these sources for errors it is interesting to
note that our results for all three states agree qualitatively with the ones in the relativistic 
quark model of Ref.~\cite{Ebert:2011jc} with quantitative deviations of at most 3~\%.

\begin{table*}[!t]
\renewcommand{\arraystretch}{1.3}
\begin{center}
\begin{tabular}{c|ccc|ccc||c|c}
\hline
\hline
            & \multicolumn{3}{c|}{$c\bar{c}   $}   & \multicolumn{3}{c||}{$b\bar{b}$}    & \multicolumn{2}{c}{$b\bar{c}$}                                         \\
$J^{PC}$    & $n=0$   & $n=1$  & $n=2$  & $n=0$   & $n=1$  & $n=2$  &  $J^{P}$ &    $n=0$                               \\
\hline                                                                                                                                                                                                     
$0^{-+}$    & $2925$  & $3684$ &        & $9414$  & $9987$ &    	&   $0^{+}$     &   $6714^{+67.1}_{-67.1}$           \\
$0^{--}$    & $3348$  &        &        & $9642$  &        &    	&   $0^{-}$   &     $6354^{+23.5}_{-23.5}$    \\
$0^{++}$    & $3323$  & $3833$ &        & $9815$  & $10254$&    	&  $1^{+}$   &                         \\
$0^{+-}$    & $3674$  &        &        & $10014$ &        &     	&  $1^{-}$  &     $6498^{+64.9}_{-64.9}$                                 \\
\hline                                                                                                                                                                                                     
$1^{-+}$    & $3524$  &        &        & $9788$  &        &    	&       &                 \\
$1^{--}$    & $3113$  & $3676$ & $3803$ & $9490$  & $10089$& $10327$&       &               \\ 
$1^{++}$    & $3489$  & $3672$ & $3912$ & $9842$  & $10120$& $10303$&       &               \\
$1^{+-}$    & $3433$  & $3747$ &        & $9806$  & $10154$&    	&       &                    \\
\hline                                                                                                                                                                                                     
$2^{-+}$    & $3806$  &        &        & $10194$  &        &    	&      &                                 \\
$2^{--}$    & $3739$  &        &        & $10145$  &        &    	&      &                                              \\
$2^{++}$    & $3550$  &        &        & $9906$   &        &    	&      &                                      \\
\hline                                                                                                                                                                                                     
$3^{--}$    & $3896$ &         &        & $10232$  &        &    	&      &                                           \\
$3^{++}$    & $3999$ &         &        & $10302$  &        &    	&       &                                     \\
$3^{+-}$    & $4037$ &         &        & $10319$  &        &    	&      &                                                   \\                                                                                                                                                                      
\hline
\hline
\end{tabular}
\caption{Calculated masses for ground and excited charmonium, bottomonium and charm-bottom states.}\label{tab:results}
\end{center}
\end{table*}

\section{Summary and conclusions}\label{sec:conclusions}

In this work we presented a first calculation of ground and excited states with angular 
momentum $J \le 3$ in the heavy quark sector using the framework of Dyson-Schwinger 
and Bethe-Salpeter equations. We have used a simple interaction model, the rainbow-ladder
approximation, which is known to represent only part of the complicated interaction
pattern of quarks and gluons even for heavy quarks. Nevertheless, we obtained surprisingly 
good results, at least for selected quantum numbers. In general, the systematics in the 
spectrum for charmonia and bottomonia is very similar, although the underlying interaction 
is not the same. Compared to the light quark sector, where
the rainbow-ladder approximation has clear deficiencies \cite{Fischer:2014xha}, the agreement with
the experimental states is much improved. This is particularly true for the ground and 
excited charmonia and bottomonia states in the vector channel, where even the second 
radial excitation is well represented. For pseudoscalar states and tensor states with quantum 
number $2^{++}$ we obtain reasonable results, whereas for scalars and axialvectors some
deviations occur. We also gave predictions for the other tensor states, in particular
for the $3^{--}$, which should be a channel where the rainbow-ladder approximation
does particularly well. For the bottomonia, our values for the tensor states may be 
considered as solid predictions for experiment with a systematic error due to 
extrapolations on the 1~\% level. We also gave results for $B_c$ states and quarkonia 
with exotic quantum numbers, although the accumulated errors in these channels due to
deficiencies in the rainbow-ladder interaction may be sizeable.

Furthermore, we studied variations of the shape of the rainbow-ladder effective coupling with 
the aim to explore whether the first or second excitation in the $1^{++}$-channel can be linked 
to the $X(3872)$ without destroying other parts of the spectrum. It turned out, that 
this is not possible, at least not within the constraints of the present study. Given the 
inherent limitation of the rainbow-ladder framework to pure vector-type interactions one
may expect that corrections beyond rainbow-ladder play an important role in this and other 
channels and may change this picture. Or it may simply be that the $X(3872)$ does 
indeed not have a strong quark-antiquark component. From the perspective of our framework, 
these questions remain open. 

Clearly, the findings of this work should be corroborated by studies beyond the 
simple rainbow-ladder scheme used in this work. Within the light meson sector
several approaches in this direction have been explored in the past 
\cite{Watson:2004kd,Fischer:2005en,Fischer:2008wy,Fischer:2009jm,Chang:2009zb,Heupel:2014ina}
and it remains to be seen whether these can be transferred to the heavy quark 
sector. This will the subject of future work.

\section*{Acknowledgments}
We thank Gernot Eichmann, Maria Gomez-Rocha, Walter Heupel, Carina Popovici 
and Helios Sanchis-Alepuz for useful discussions. 
This work was supported by the BMBF under contract No. 06GI7121, 
the Helmholtz International Center for FAIR within the LOEWE program of the 
State of Hesse, and the Austrian Science Fund (FWF) under project number M1333-N16.

\end{document}